\documentclass[aps,prl,twocolumn,superscriptaddress,preprintnumbers,amsmath,amssymb,longbibliography,10pt]{revtex4-2}
\usepackage{dcolumn}
\usepackage{graphicx}
\usepackage{amsmath,amsfonts,amssymb,amsthm,amstext,amsbsy,amsopn,amscd,amsxtra}
\usepackage{dsfont}
\usepackage{physics} 
\usepackage{upgreek} 
\usepackage[colorlinks=true,allcolors=blue]{hyperref}
\usepackage{xcolor}
\usepackage{url}
\usepackage{siunitx}
\usepackage{multirow}
\usepackage{hhline}
\usepackage{bm}

\everymath{\displaystyle}

\begin{document}

\title{Control of the Purcell effect via unexcited atoms and exceptional points}

\author{G. S. Agarwal}
\email{Girish.Agarwal@ag.tamu.edu}
\affiliation{Institute for Quantum Science and Engineering, Department of Physics and Astronomy, Department of Biological and Agricultural Engineering, Texas A\&M University, College Station, Texas 77843, USA}

\date{\today}

\begin{abstract}

We examine the possible control of the celebrated Purcell effect in cavity quantum electrodynamics. We demonstrate that the presence of an unexcited atom can significantly alter the Purcell decay depending on the strength of coupling of the unexcited atom with the cavity mode though the excited atom has to be weakly coupled for it to be in the Purcell regime. This is distinct from the nonradiative nature of the singlet state which is an entangled state of the two atom system. We present physical interpretation for inhibition as due to interference between two polariton channels of decay. We bring out connection to exceptional points in the cavity QED system as the unexcited atom and cavity mode can produce a second order exceptional point. We further show how two unexcited atoms can create a third order exceptional point leading to inhibition of Purcell effect. We also discuss the case when the Purcell effect can be enhanced.
\end{abstract}

\maketitle

{\it Introduction.}--
The Purcell effect has been the hallmark of cavity quantum electrodynamics \cite{Purcell1995,goy1983observation,angelakis2004photonic,kockum2018decoherence,collison2021purcell,krishnamoorthy2012topological,krasnok2015antenna,agarwal1975quantum}. It brought for the first time the role of coupling to the cavity and how this coupling can lead to the decay of the atom even if the decay of the atom otherwise is negligible. The enhanced decay results due to enhancement of the density of states in the cavity. Analogues of similar enhancement in the density of states and the resulting Purcell effect are known in other contexts, for example in the decay of the atom located in the close vicinity of the metal \cite{collison2021purcell,krishnamoorthy2012topological,agarwal1975quantum} and in other photonic structures \cite{kockum2018decoherence,rao2007single}. Cavities thus provide an effective way to deal with the manipulation of spontaneous emission and these can also lead to inhibition of spontaneous emission \cite{angelakis2004photonic,kleppner1981inhibited} if the density of states at the location of the atom is negligible. Other methods to control spontaneous emission are based on photonic crystals \cite{yablonovitch1987inhibited,john1990quantum,leistikow2011inhibited}, whispering gallery modes \cite{he2011detecting} etc; one dimensional photonic waveguides \cite{kockum2018decoherence}. A new direction in recent years, is to use exceptional points [EP] for studying various effects like enhanced sensing capabilities of parameters when perturbed in the vicinity of the EP \cite{wiersig2014enhancing,chen2017exceptional,hodaei2017enhanced,Wiersig2020,lau2018fundamental,chen2019sensitivity,anderson2023clarification}. Such enhanced capabilities have been demonstrated especially in systems with PT symmetry i.e. with both gain and loss \cite{el2018non,ozdemir2019parity}. Enhanced sensing at both second order \cite{chen2017exceptional} and higher order EP's \cite{hodaei2017enhanced} have been studied. While these are semiclassical studies and there exist plenty of such studies; the QED effects near EP's are still scant \cite{khanbekyan2020decay}. Early studies focused on spontaneous generation of photons \cite{agarwal2012spontaneous}, where as more recent ones examined the effects of quantum fluctuations on signal to noise ratio in sensing \cite{lau2018fundamental,chen2019sensitivity,anderson2023clarification}. Other applications of EP in the context of quantum physics are starting to appear \cite{li2023speeding,khanbekyan2020decay,mukamel2023exceptional}. A recent one presents speed up of the generation of quantum entanglement near EP \cite{li2023speeding}.

In this work we consider how the presence of an unexcited atom can influence significantly the Purcell effect of an excited atom. The two atoms are differently coupled to the cavity. The excited atom is weakly coupled to the cavity and this can be achieved by placing it close to the node of the cavity mode. We consider several different scenarios for strong inhibition -- The first scenario is such that the atom's coupling leads to large cooperativity parameter. In other scenarios we can use unexcited atoms to create EP's of second and higher orders. It is easy to create second order EP whereas it is more involved to create higher order EP's. We create third order EP by choosing two unexcited atoms with appropriate decay parameters and then examine the decay of a weakly coupled third atom. In each case we present approximate analytical results for the inhibition of the Purcell decay. We also present physical interpretation of the inhibition in terms of the cavity polaritons \cite{hopfield1958theory}. The approximate results are confirmed from the solution of full quantum master equation describing the dynamics of atoms in a cavity. The inhibition effects are especially pronounced for Rydberg atoms which were used in the first study of the Purcell effect \cite{goy1983observation}. For detuned excited atom, the Purcell decay can be enhanced by the unexcited atom. In this case, the enhancement arises as the detuned excited atom can come in resonance with one of the cavity polariton states. Thus in general, the unexcited atom can lead to the significant inhibition or enhancement of the Purcell decay. The results that we report are quite generic to any cavity QED system and can be studied experimentally for example with superconducting qubits in cavities \cite{mlynek2014observation} or with atoms trapped in cavities \cite{liu2023realization}.

\begin{figure}[b]
 	\includegraphics[width=8.6cm]{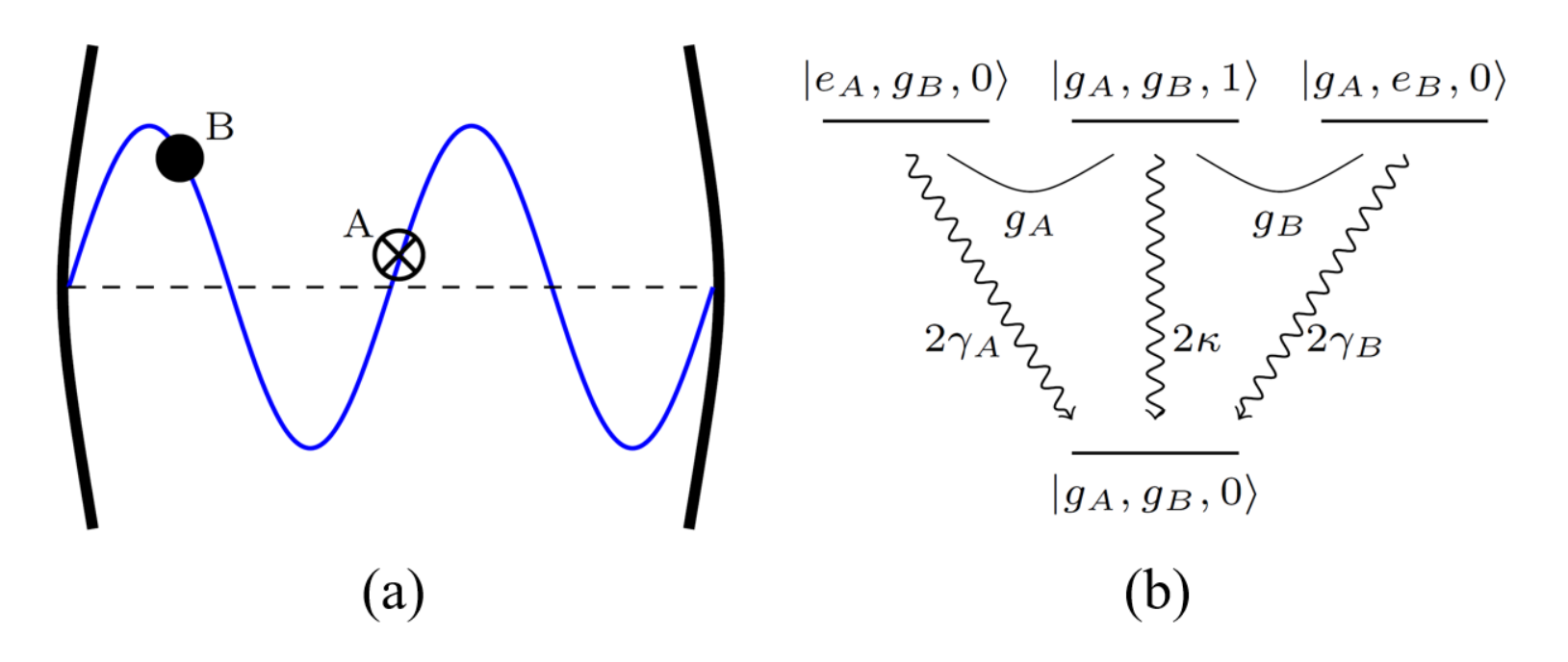}
 	\caption{Purcell decay of an excited atom A in presence of an unexcited atom B and the relevant quantum states participating in dynamical evolution.}
 	\label{fig1}
 \end{figure}

{\it Control of the Purcell Decay.}--
 In order to demonstrate how the Purcell decay can be controlled, we start with the simplest model [Fig. \ref{fig1}(a)] with two atoms in a cavity with frequency $\omega_c$. For simplicity we assume that both atoms are on resonance with the cavity mode. Let $2\kappa$ be the leakage rate of photons from the cavity and let $g_A$ be the coupling constant of the excited atom A to the cavity mode. Then the Purcell decay of the excited atom is given by $2\Gamma=2g_A^2/\kappa$ if $g_A^2/\kappa^2\ll1$. Let us now introduce an unexcited atom B with a coupling constant $g_B$ with the cavity mode. We will keep $g_B$ flexible -- we will examine both cases when $g_B$ is large and when $g_B$ is small. This can be achieved by moving the position of the atom B along the cavity mode function. The interaction Hamiltonian in a frame rotating with the frequency $\omega_c$ is given by
 \begin{equation}
 	H=\hbar g_A(a^\dagger S_A^-+aS_A^+)+\hbar g_B(a^\dagger S_B^-+aS_B^+),
 \end{equation}
 where $S_A^+$ and $S_B^+$ are the excitation operators for atoms A and B respectively. The master equation of the atoms and the cavity mode is
 \begin{equation}
 	\begin{aligned}
 		\frac{\partial\rho}{\partial t}&=-\frac{i}{\hbar}[H,\rho]-\kappa(a^\dagger a\rho-2a\rho a^\dagger+\rho a^\dagger a)\\
 		&\qquad-\gamma_A(S_A^+S_A^-\rho-2S_A^-\rho S_A^++\rho S_A^+S_A^-)\\
 		&\qquad-\gamma_B(S_B^+S_B^-\rho-2S_B^-\rho S_B^++\rho S_B^+S_B^-),
 	\end{aligned}
 	\label{master}
 \end{equation}
 where atom A [B] decays from the excited state $|e\rangle$ to the ground state $|g\rangle$ at the rate $2\gamma_A$ [$2\gamma_B$]. The initial state of the system consisting of two atoms and cavity mode is $|e_A,g_B,0\rangle$, i.e., atom A is excited and both atom B and the cavity mode are in ground state. Since there is only one excitation in the system, during dynamical evolution we have the following possibilities as illustrated in Fig. \ref{fig1}(b): 
 $|\psi_1\rangle=|e_A,g_B,0\rangle\stackrel{g_A}\rightleftarrows|g_A,g_B,1\rangle=|\psi_2\rangle$; $|g_A,g_B,1\rangle\stackrel{g_B}\rightleftarrows|g_A,e_B,0\rangle=|\psi_3\rangle$; $|e_A,g_B,0\rangle\stackrel{\gamma_A}\rightarrow|g_A,g_B,0\rangle=|\psi_4\rangle$; $|g_A,g_B,1\rangle\stackrel{\kappa}\rightarrow|g_A,g_B,0\rangle$; $|g_A,e_B,0\rangle\stackrel{\gamma_B}\rightarrow|g_A,g_B,0\rangle$. Thus the time evolution is restricted to only four states $|\psi_j\rangle$, $j=1,\ldots,4$. It is to be noted that no population is transferred back from the state $|\psi_4\rangle$ to the states $|\psi_i\rangle$, $i=1,2,3$. This observation allows us to express elements of the density matrix as $(\rho(t))_{ij}=\psi_i(t)\psi_j^*(t)$, $i=1,2,3$, $j=1,2,3$. This procedure is exact. Using Eq. (\ref{master}), the time evolution is then given by $\dot\psi_i(t)=-iM\psi$, with
 \begin{equation}
 	M=\left(\begin{array}{ccc}
 		-i\gamma_A & g_A & 0\\
 		g_A & -i\kappa & g_B\\
 		0 & g_B & -i\gamma_B
 	\end{array}
 	\right).
 	\label{MM}
 \end{equation}
 The eigenvalues $\lambda$ of $M$ are given by
 \begin{equation}
 	(\lambda+i\kappa)(\lambda+i\gamma_A)(\lambda+i\gamma_B)-g_B^2(\lambda+i\gamma_A)-g_A^2(\lambda+i\gamma_B)=0.
 	\label{chara}
 \end{equation}
 To see the emergence of the Purcell effect, we check the relevant eigenvalues for $g_B=0$, $\gamma_A=0$. These are $-\frac{i\kappa}{2}\pm\frac{i}{2}\sqrt{\kappa^2-4g_A^2}\rightarrow-i\kappa;-\frac{ig_A^2}{\kappa}$ in the limit of $\frac{4g^2}{\kappa^2}\ll1$. The eigenvalue $-i\kappa$ is the cavity decay whereas the eigenvalue $-ig_A^2/\kappa$ gives the Purcell decay. In presence of the unexcited atom B, $g_B\neq0$, we can evaluate perturbativity the Purcell decay as $4g_A^2/\kappa^2$ is small. Note that for $g_A=0$, the two eigenvalue are given by
 \begin{equation}
 	\lambda_\pm=-\frac{i(\kappa+\gamma_B)}{2}\pm\sqrt{g_B^2-\left(\frac{\kappa-\gamma_B}{2}\right)^2},
 	\label{eig}
 \end{equation}
 which give the energies and decays of the well known dressed states, also known as cavity polariton states. The eigenvalues associated with Eq. (\ref{chara}) will have a perturbative correction term of order of $g_A^2$. The perturbative corrections to Eq. (\ref{eig}) can be ignored because of the large term $-i(\kappa+\gamma_B)/2$. The other eigenvalue which was $g_A^2/\kappa$ if $g_B=0$, now becomes
 \begin{equation}
 	\tilde\Gamma=\Gamma/(1+\frac{g_B^2}{\gamma_B\kappa})=\Gamma/(1+C),\qquad \Gamma=\frac{g_A^2}{\kappa},
 	\label{inhi}
 \end{equation}
 where $C$ is the coorperativity parameter $g_B^2/\gamma_B\kappa$. We thus find that the Purcell decay constant of the excited state of the atom A can be inhibited by a factor $1+C$, i.e., by a large amount if the cooperativity parameter $C$ is large. For strong coupling of the unexcited atom, the inhibition is quite large. In such a situation the population remains trapped in excited state! We remind the reader that this is different from subradiance which requires two atoms to be prepared in a singlet state which is an entangled state, in addition to the requirement that the two atoms have identical coupling to the cavity. In contrast for our case $g_A\neq g_B$ and that the initial state is an unentangled state $|e_A,g_B,0\rangle$. While we have derived the result given by Eq. (\ref{inhi}) from the consideration of the eigenvalues of the matrix $M$ which gives the evolution of populations in the excited states, we have also confirmed Eq. (\ref{inhi}) from the full time dependent solution of the master equation (\ref{master}). We evaluated the population in $|e_A,g_B,0\rangle$, i.e., $\rho_{11}(t)$ and fit it to an exponential function $\exp(-2\tilde\Gamma t)$. This fitting yields the result Eq. (\ref{inhi}) -- for instance, $\tilde\Gamma/\Gamma=1$, $0.1$, $0.04$ for $g_B/\kappa=0$, $3$ and $5$ respectively and for $\gamma_B/\kappa=1$. We also note in passing that Eq. (\ref{chara}) admits other behavior as well. For example, for identical atoms, if $\gamma_A=\gamma_B\sim0$, $g_A=g_B=g$, then eigenvalues are $\lambda=0$, $\lambda(\lambda+i\kappa)-2g^2=0$, which lead to subradiant and superradiant character \cite{mlynek2014observation}.

{\it Purcell Decay at the second ord EP.}-- 
\begin{figure}[b]
	\includegraphics[width=6.6cm]{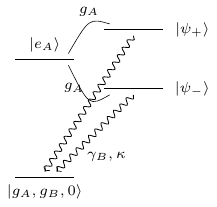}
	\caption{Decay channels for the excited state of atom A via the cavity polaritons/dressed states $|\psi_\pm\rangle$ with separation $2\sqrt{g_B^2-(\frac{\kappa-\lambda}{2})^2}$ if $g_B>\left|\frac{\kappa-\lambda}{2}\right|$.}
	\label{fig2}
\end{figure}
Note that the cavity QED system can produce an EP if $g_B=|\frac{\kappa-\gamma_B}{2}|$, then from Eq. (\ref{eig}) the two degenerate eigenvalues are $-\frac{i(\kappa+\gamma_B)}{2}$ and the two eigenfunctions coalesce into $(1,i{\rm sign}(\kappa-\gamma_B))^T/\sqrt2$, where $T$ stands for transpose. This is the second order EP which is different from what happens in PT symmetric systems where the two eigenvalues are strictly zero. At the second order EP created by the unexcited atom B, the result Eq. (\ref{inhi}) becomes
\begin{equation}
	\tilde\Gamma=\Gamma(1-(\frac{\kappa-\gamma_B}{\kappa+\gamma_B})^2),
	\label{sec}
\end{equation}
which can become much smaller than $\Gamma$ if $\kappa/\gamma_B\ll1$ or $\kappa/\gamma_B\gg1$. Again the validity of the result (\ref{sec}) has been confirmed from the solution of the master equation (\ref{master}). For instance $\tilde\Gamma/\Gamma=0.555$ for $\gamma_B/\kappa=5$ or $\kappa/\gamma_B=5$ and with $g_B$ given in terms of $\gamma_B$ and $\kappa$, i.e., $g_B=\frac{|\gamma_B-\kappa|}{2}$. 

The inhibition can also be understood in terms of the dressed states $|\psi_{\pm}\rangle$ or cavity polaritons created by the atom B in the absence of atom A(see Fig. \ref{fig2}). These correspond to the eigenvalues $\lambda_\pm$ (Eq. (\ref{eig})). The states $|\psi_\pm\rangle$ are linear combinations of the states $|e_B,0\rangle$ and $|g_B,1\rangle$, i.e., $(|g_B,1\rangle\pm|e_B,0\rangle)/\sqrt2$, where $|e_B\rangle$ and $|g_B\rangle$ are the excited and ground states of atom B. The excited states $|e_A\rangle$ couples to both $|\psi_\pm\rangle$ via the component $|g_B,1\rangle$. Thus in the Purcell decay of atom A, there will be interference effects between the two transition amplitudes corresponding to channels $|e_A\rangle\stackrel{g_A}\rightarrow|\psi_\pm\rangle|g_A\rangle\stackrel{\gamma_B,\kappa}\rightarrow|g_A,g_B,0\rangle$ leading to the inhibition of decay of atom A.

We like to add that spontaneous emission at EP has been considered before for two different models -- Agarwal \cite{agarwal2023laser} used a {\it single} Lambda atomic scheme to create double pole in S matrix via coupling to a laser field and then considered spontaneous emission under weak coupling on the transition uncoupled to the laser field \cite{harris1990nonlinear,liang2023observation}. Khanbekyan and Wiersig \cite{khanbekyan2020decay} considered suppression of spontaneous emission from an atom coupled strongly to a high quality double mode cavity. The two modes of the cavity created a second order pole in the S matrix. The model that we present is quite different we use an unexcited atom or even an unexcited ensemble to inhibit Purcell effect. Our work is based on coupling to a single mode of the cavity. This is quite counterintuitive. In our model the two atoms are differently coupled. Thus standard subradiant effects \cite{scully2015single} are not the reason for the inhibition of the Purcell effect as the conditions for subradiant effects are rigid and require preparation of the entangled state of the two atoms coupled equally to the cavity mode.

{\it Purcell decay at the third order EP.}-- 
\begin{figure}[b]
	\includegraphics[width=6.6cm]{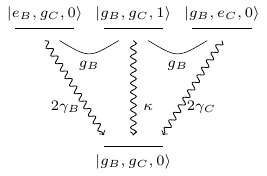}
	\caption{Creation of the third order EP in cavity QED under conditions given by Eq. (\ref{cond}) with two atoms coupled to a cavity mode via the single excitation manifold.}
	\label{fig3}
\end{figure}
We will first discuss how a third order EP can be created in cavity QED. We will now use two unexcited atoms B and C coupled similarly to the cavity mode. Let us first assume that the excited atom A is absent, but assume one photon in the cavity. This is to bring out the dressed state structure and the third order exception point in the cavity QED system. In terms of the states $|e_B,g_C,0\rangle$, $|g_B,g_C,1\rangle$ and $|g_B,e_C,0\rangle$, the matrix giving time evolution, of the system consisting of atoms B, C and the cavity mode, would be similar to Eq. (\ref{MM})
\begin{equation}
	\begin{aligned}
		M_0&=\left(\begin{array}{ccc}
			-i(\gamma_B-\kappa) & g_B & 0\\
			g_B & 0 & g_B\\
			0 & g_B & -i(\gamma_C-\kappa)
		\end{array}
		\right)-i\kappa\\
		&=\left(\begin{array}{ccc}
			i\tilde\gamma & g_B & 0\\
			g_B & 0 & g_B\\
			0 & g_B & -i\tilde\gamma
		\end{array}
		\right)-i\kappa,
	\end{aligned}
	\label{4mm}
\end{equation}
where we choose decays $\gamma_B$ and $\gamma_C$ so that matrix in Eq. (\ref{4mm}) has structure similar to PT symmetric systems \cite{hodaei2017enhanced},
\begin{equation}
	\tilde\gamma=\gamma_C-\kappa,\qquad \gamma_B=2\kappa-\gamma_C,\qquad \frac{\gamma_C}{2}<\kappa<\gamma_C.
	\label{cond}
\end{equation}
The eigenvalues of $M_0$ are given by 
\begin{equation}
	\lambda=-i\kappa,\qquad -i\kappa\pm\sqrt2g_B\sqrt{1-(\frac{\tilde\gamma}{\sqrt2g_B})^2},
\end{equation}
the corresponding eigenfunctions of $M_0$ will give three polariton states. For $\tilde\gamma=\sqrt2g_B$, we have a third order EP which is created by two atoms interacting with the cavity mode as shown in Fig. \ref{fig3}. We need to choose two atoms with different decay constants $\gamma_B\neq\gamma_C$, so that conditions in Eq. (\ref{cond}) are satisfied. Note that all three polaritons have the same decay constant $\kappa$ with energies given by $0$, $\pm\sqrt2g_B\sqrt{1-(\tilde\gamma/\sqrt2g_B)^2}$. These coalesce into degenerate states if $\tilde\gamma=\sqrt2 g_B$.

We can now consider Purcell decay of an excited atom A in presence of two unexcited atoms B and C. We start with the generalization of the master equation (\ref{master}) and consider the relevant excited states $|e_A,g_B,g_C,0\rangle$, $|g_A,e_B,g_C,0\rangle$, $|g_A,g_B,e_C,0\rangle$, $|g_A,g_B,g_C,1\rangle$. Then instead of the dynamical matrix given by Eq. (\ref{MM}), we get the $4\times4$ matrix
\begin{equation}
	M=\left(\begin{array}{cccc}
		0 & 0 & g_A & 0\\
		0 & -i\gamma_B & g_B & 0\\
		g_A & g_B & -i\kappa & g_B\\
		0 & 0 & g_B& -i\gamma_C
	\end{array}
	\right).
	\label{44m}
\end{equation}
Following the same procedure as employed in connection with Eq. (\ref{MM}), we find that
\begin{equation}
	\bar\Gamma=\Gamma/[1+\frac{2g_B^2}{\gamma_C(2\kappa-\gamma_C)}],
\end{equation}
which at third order EP simplifies to
\begin{equation}
	\tilde\Gamma=\Gamma(\frac{\gamma_C}{\kappa})(2-\frac{\gamma_C}{\kappa})=\Gamma(\frac{\gamma_B}{\kappa})(2-\frac{\gamma_B}{\kappa}),\qquad \frac{\gamma_C}{2}<\kappa<\gamma_C.
\end{equation}
Thus at third order EP, we can again have significant inhibition of Purcell emission. For instance for $\gamma_C/\kappa=1.95$, $\tilde\Gamma/\Gamma=0.0975$. These results can be confirmed from the full solution of the master equation for three atoms interacting with the cavity mode. In terms of the cavity polaritons, we have a picture similar to Fig. \ref{fig2}, i.e., the excited atom has two channels of decay: $|e_A\rangle\rightarrow|\psi_\pm\rangle|g_A\rangle\rightarrow|g_A,g_B,g_C,0\rangle$; $|\psi_\pm\rangle=\frac12(|e_B,g_C,0\rangle+|g_B,e_C,0\rangle\pm\sqrt2|g_B,g_C,1\rangle)$. The excited atom does not couple to the other polariton state $|\psi_0\rangle=\frac1{\sqrt2}(|e_B,g_C,0\rangle-|g_B,e_C,0\rangle)$ as it has no cavity photon component.

{\it Enhanced Purcell decay by an unexcited atom.}-- 
\begin{figure}[b]
	\includegraphics[width=8.6cm]{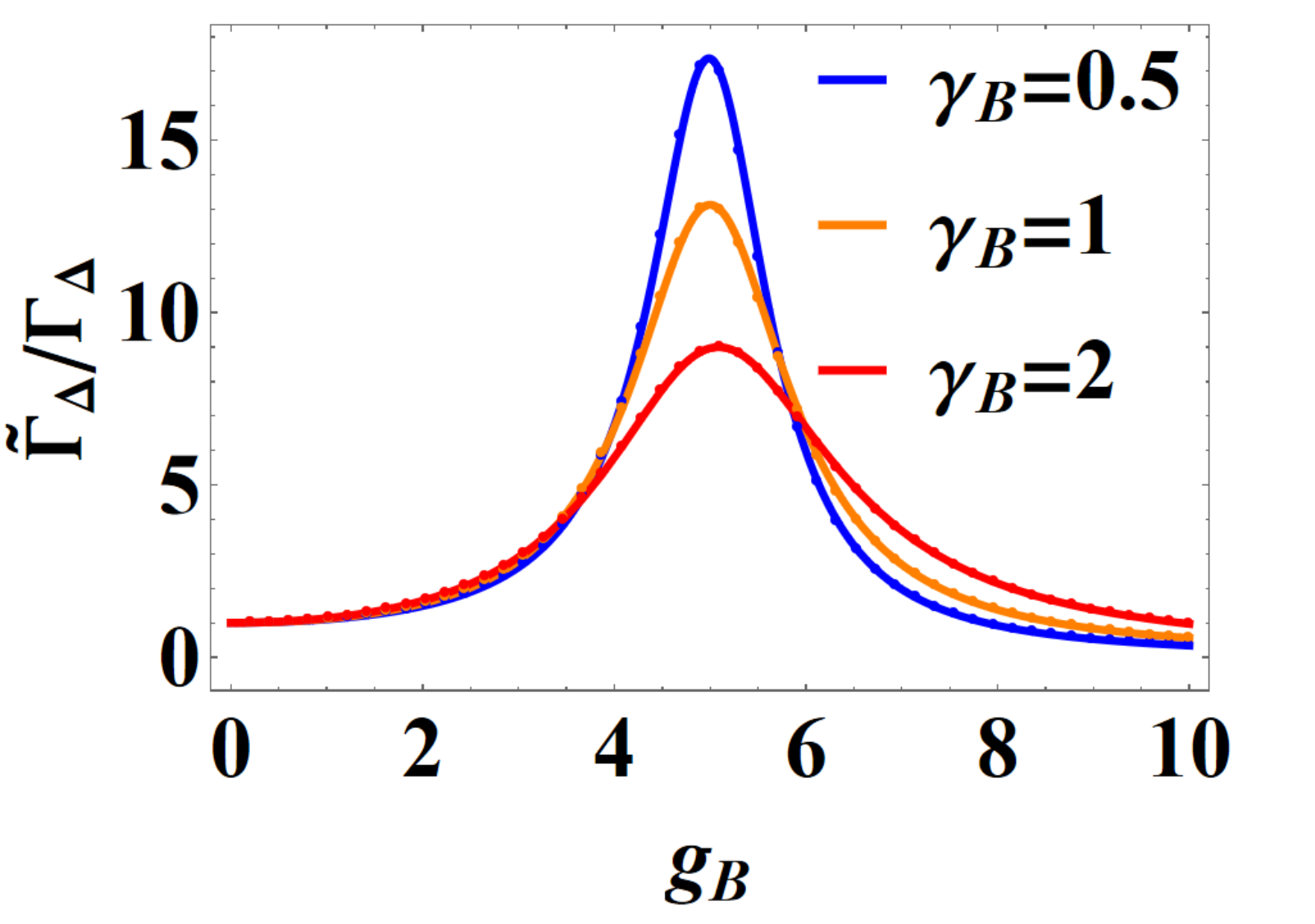}
	\caption{Enhancement of the Purcell decay in presence of an unexcited atom coupled strongly to the cavity, $\Delta=5$, with all frequencies in units of $\kappa$.}
	\label{fig4}
\end{figure}
We now demonstrate how the Purcell effect can be enhanced due to the presence of an unexcited atom B. We will assume that the atom B is on resonance with the cavity whereas the atom A is detuned by an amount $\Delta$. The well known Purcell decay rate when $\Delta\neq0$, is
\begin{equation}
	\Gamma_\Delta=g_A^2\kappa/(\kappa^2+\Delta^2),
	\label{gmdt}
\end{equation}
which is negligible if $\Delta/\kappa$ is large. 

We can now repeat the analysis that led to Eq. (\ref{inhi}) to obtain modification of Eq. (\ref{gmdt}) due to the presence of an unexcited atom. The eigenvalues of the matrix describing the evolution of the states $|\psi_i\rangle$ given before Eq. (\ref{44m}) are given by ($\gamma_A=0$)
\begin{equation}
	(\lambda+i\kappa)(\lambda-\Delta)(\lambda+i\gamma_B)-g_B^2(\lambda-\Delta)-g_A^2(\lambda+i\gamma_B)=0.
	\label{lbkp}
\end{equation}
An analysis of Eq. (\ref{lbkp}) shows that
\begin{equation}
	\begin{aligned}
		\tilde\Gamma_\Delta&=g_A^2{\rm Im}\{\frac{g_B^2}{\Delta+i\gamma_B}-i\kappa-\Delta\}^{-1}\\
		&=-g_A^2{\rm Im}\{\frac{\Delta+i\gamma_B}{(\Delta-\lambda_+)(\Delta-\lambda_-)}\},
	\end{aligned}
\end{equation}
where $\lambda_\pm$ are given by Eq. (\ref{eig}), i.e., these are the complex energies of the two polariton states formed by the interaction of atom B with the cavity. For large $\Delta$, comparable to the polariton energy ${\rm Re}\lambda_+$, $\tilde\Gamma_\Delta$ can be significantly enhanced over $\Gamma_\Delta$. This is confirmed by the plot in Fig. \ref{fig4}. The actual amount of enhancement depends on the decay parameter $\gamma_B$ of the unexcited atom. We note that with $\Delta\neq0$, the density of states available for Purcell decay is less than for $\Delta=0$. With the presence of the atom B, the atom A can interact resonantly with the polariton states $|\psi_\pm\rangle$ leading to much higher density of states and hence to enhanced emission.

{\it Conclusions.}-- 
We have demonstrated how the Purcell decay of an atom coupled weakly to a cavity can be controlled by the presence of unexcited atoms. We have presented various physical mechanisms for the inhibition or enhancement of the Purcell decay. We bring out the role of the polariton states created by unexcited atoms and the coupling of these polariton states to the excited atom. These polariton states provide several decay channels to the excited atom and depending on the strength of coupling of the unexcited atoms to the cavity, there could be interferences among the decay channels. We also show that the unexcited atoms can create exceptional points in the cavity QED set up. The decay of the excited atom can then be related to decay at second and higher order exceptional points. The effects that we report are different from the more traditional superradiant and subradiant effects which require entangled states. 

{\it Acknowledgments.}--
We thank the Air Force Office of Scientific Research (Award No. FA-9550-20-1-0366) and the Robert A. Welch Foundation (Grant No. A-1943-20210327) for supporting this work. We thank Q. Miao for help in preparing this manuscript.

\bibliography{main}

\end{document}